\documentclass[
   ,final            
  ]
  {aipproc}

\layoutstyle{8x11single}

\begin{document}

\title{Nucleon Resonance Excitation with Virtual Photons}

\classification{13.40.Gp, 14.20.Gk, 11.80.Et, 13.30.Eg}

\keywords{Partial-Wave Analysis, Pion Electroproduction, Transition Form Factors,
Helicity Amplitudes }

\author{L.~Tiator}{
  address={Institut für Kernphysik, Universität Mainz, 55099 Mainz, Germany}
  }

\author{S.~Kamalov}{
  address={JINR Dubna, 141980 Moscow Region, Russia}
  }

\begin{abstract}
The unitary isobar model MAID is used for a partial wave analysis of pion photoproduction
and electroproduction data on the nucleon. In particular we have taken emphasis on the
region of the $\Delta(1232)$ resonance and have separated the resonance and background
amplitudes with the K-matrix approach. This leads to electromagnetic properties of the
dressed $\Delta$ resonance, where all multipole amplitudes become purely imaginary and
all form factors and helicity amplitudes become purely real at the K-matrix pole of
$W=M_\Delta=1232$~MeV. The $R_{EM}=E2/M1$ and $R_{SM}=C2/M1$ ratios of the quadrupole
excitation are compared to recent data analysis of different groups. The $R_{EM}$ ratio
of MAID2005 agrees very well with the data and has a linear behavior over the whole
experimentally explored $Q^2$ region with a small positive slope that predicts a zero
crossing around $3.5$~GeV$^2$. The recent $R_{SM}$ data for $Q^2 < 0.2$~GeV$^2$ indicate
a qualitative change in the shape of the ratio which can be explained by the impact of
the Siegert theorem at pseudothreshold ($Q^2=-0.086$GeV$^2)$ in the unphysical region.
\end{abstract}

\maketitle

\section{Introduction}
\label{sec1}

Our knowledge of nucleon resonances is mostly given by elastic pion nucleon scattering
\cite{Hohler79}. All resonances that are given in the Particle Data Tables \cite{PDG04}
have been identified in partial wave analyses of $\pi N$ scattering both with
Breit-Wigner analyses and with speed-plot techniques. From these analyses we know very
well the masses, widths and the branching ratios into the $\pi N$ and $\pi\pi N$
channels. These are reliable parameters for all resonances in the 3- and 4-star
categories. There remain some doubts for the two prominent resonances, the Roper
$P_{11}(1440)$ which appears unusually broad and the $S_{11}(1535)$ that cannot uniquely
be determined in the speed-plot due to its position close to the $\eta N$ threshold. As
an alternative to the quark model, these resonances can also be generated with dynamical
methods \cite{Sieg95,Krew00,Lutz04}. On the other hand attempts have been made to
calculate nucleon resonances in quenched QCD on the lattice~\cite{Lee03}, and in the case
of the $\Delta(1232)$ even the transition form factors are evaluated on the
lattice~\cite{Alex05}.

Starting from these firm grounds, using pion photo- and electroproduction we can
determine the electromagnetic $\gamma N N^*$ couplings. They can be given in terms of
electric, magnetic and charge transition form factors $G_E^*(Q^2)$, $G_M^*(Q^2)$ and
$G_C^*(Q^2)$ or by linear combinations as helicity amplitudes $A_{1/2}(Q^2)$,
$A_{3/2}(Q^2)$ and $S_{1/2}(Q^2)$. So far, we have some reasonable knowledge of the
$A_{1/2}$ and $A_{3/2}$ amplitudes at $Q^2=0$, which are tabulated in the Particle Data
Tables. For finite $Q^2$ the information found in the literature is very scarce and
practically does not exist at all for the longitudinal amplitudes $S_{1/2}$. But even for
the transverse amplitudes only few results are firm, these are the $G_M^*$ form factor of
the $\Delta(1232)$ up to $Q^2\approx 10$~GeV$^2$, the $A_{1/2}(Q^2)$ of the
$S_{11}(1535)$ resonance up to $Q^2\approx 5$~GeV$^2$ and the helicity asymmetry
$A_1(Q^2)$ for the $D_{13}(1520)$ and $F_{15}(1680)$ resonance excitation up to
$Q^2\approx 3$~GeV$^2$ which change rapidly between $-1$ and $+1$ at small $Q^2\approx
0.5$~GeV$^2$ \cite{Boffi96}. Frequently also data points for other resonance amplitudes,
e.g. for the Roper are shown together with quark model calculations but they are not very
reliable. Their statistical errors are often quite large but in most cases the model
dependence is as large as the absolute value of the data points. In this context it is
worth noting that also the word `data point` is somewhat misleading because these photon
couplings and amplitudes cannot be measured directly but can only be derived in a partial
wave analysis. Only in the case of the $\Delta(1232)$ resonance this can and has been
done directly in the experiment by Beck et al. at Mainz \cite{Beck97}. For the Delta it
becomes possible due to two important theoretical facts, the Watson theorem and the well
confirmed validity of the $s$+$p$ -- wave truncation. Within this assumption a complete
experiment was done with polarized photons and with the measurement of both $\pi^0$ and
$\pi^+$ in the final state, allowing also for an isospin separation. For other resonances
neither the theoretical constraints are still valid nor are we any close to a complete
experiment. The old data base was rather limited with large error bars and no data with
either target or recoil polarization was available. Even now we do not have many data
points with double polarization, however, the situation for unpolarized $e+p\rightarrow
e'+p+\pi^0$ has considerably improved, mainly by the new JLab experiments in all three
halls A, B and C. Furthermore electron beam polarization has been used in a couple of
experiments at Mainz, Bates and JLab. These data cover a large energy range from the
Delta up to the third resonance region with a wide angular range in $\theta_\pi$. Due to
the $2\pi$ coverage in the $\phi$ angle, a separation of the unpolarized cross section
\begin{equation}
\frac{d\sigma_v}{d\Omega}= \frac{d\sigma_T}{d\Omega} +
\varepsilon\frac{d\sigma_L}{d\Omega}
+\sqrt{2\varepsilon(1+\varepsilon)}\frac{d\sigma_{LT}}{d\Omega}\cos\phi
+\varepsilon\frac{d\sigma_{TT}}{d\Omega}\cos 2\phi
+h\sqrt{2\varepsilon(1-\varepsilon)}\frac{d\sigma_{LT^\prime}}{d\Omega}\sin\phi
\label{eq1}
\end{equation}
in four parts becomes possible and is very helpful for the partial wave analysis. Even
without a Rosenbluth separation of $d\sigma_T$ and $d\sigma_L$ we have an enhanced
sensitivity of the longitudinal amplitudes due to the $d\sigma_{LT}$ and $d\sigma_{LT'}$
interference terms. Such data are the basis of our new partial wave analysis with an
improved version of the Mainz unitary isobar model MAID.

\section{The dynamical approach to meson electroproduction}
\label{sec2}

In the dynamical approach to pion photo- and electroproduction~\cite{Yang85,KY99,DMT01},
the t-matrix is expressed as
\begin{eqnarray}
t_{\gamma\pi}(E)=v_{\gamma\pi}+v_{\gamma\pi}\,g_0(E)\,t_{\pi N}(E)\,, \label{eq2}
\end{eqnarray}
where $v_{\gamma\pi}$ is the transition potential for $\gamma^*N \rightarrow \pi N$, and
$t_{\pi N}$ and $g_0$ denote the $\pi N$ t-matrix and free propagator, respectively, with
$E \equiv W$ the total energy in the c.m. frame. A multipole decomposition of
Eq.~(\ref{eq2}) gives the physical amplitude
\begin{eqnarray}
&&t_{\gamma\pi}^{\alpha}(q,k;E+i\epsilon)
=\exp{(i\delta^{\alpha})}\,\cos{\delta^{\alpha}} \left[ v_{\gamma\pi}^{\alpha}(q,k)
+P\int_0^{\infty} dq' \frac{q'^2R_{\pi
N}^{\alpha}(q,q';E)\,v_{\gamma\pi}^{\alpha}(q',k)}{E-E_{\pi N}(q')}\right]\,, \label{eq3}
\end{eqnarray}
where $\delta^{\alpha}$ and $R_{\pi N}^{\alpha}$ are the $\pi N$ scattering phase shift
and reaction matrix in channel $\alpha$, respectively; $q$ is the pion on-shell momentum
and $k$ is the virtual photon momentum for a four-momentum transfer of $-Q^2$,
\[
q=\frac{\sqrt{(W^2-(M_N+m_\pi)^2)(W^2-(M_N-m_\pi)^2)}}{2W},\quad
k=\frac{\sqrt{(Q^2+(W+M_N)^2)(Q^2+(W-M_N)^2)}}{2W}.
\]
For simplicity we give here the expressions which are strictly valid only in the limit of
the Watson theorem. They are exact for the $\Delta(1232)$ but need to be modified for
other resonances as discussed in detail in Ref.~\cite{Chen03}.

The multipole amplitude in Eq.~(\ref{eq3}) manifestly satisfies the Watson theorem and
shows that the $\gamma,\pi$ multipoles depend on the half-off-shell behavior of the $\pi
N$ interaction.

In a resonant channel the transition potential $v_{\gamma\pi}^\alpha$ consists of two
terms
\begin{eqnarray}
v_{\gamma\pi}^\alpha(W,Q^2)=v_{\gamma\pi}^{B,\alpha}(W,Q^2)
+v_{\gamma\pi}^{R,\alpha}(W,Q^2),\label{eq4}
\end{eqnarray}
where $v_{\gamma\pi}^{B,\alpha}(W,Q^2)$ is the background transition potential and
$v_{\gamma\pi}^{R,\alpha}(W,Q^2)$ corresponds to the contribution of the bare resonance
excitation. The resulting t-matrix can be decomposed into two terms
\begin{eqnarray}
t_{\gamma\pi}^\alpha(W,Q^2)=t_{\gamma\pi}^{B,\alpha}(W,Q^2) +
t_{\gamma\pi}^{R,\alpha}(W,Q^2)\,.\label{eq5}
\end{eqnarray}

The background  potential $v_{\gamma\pi}^{B,\alpha}(W,Q^2)$ is described by Born terms
obtained with an energy dependent mixing of pseudovector-pseudoscalar $\pi NN$ coupling
and t-channel vector meson exchanges. The mixing para\-meters and coupling constants were
determined from an analysis of nonresonant multipoles in the appropriate energy regions.
In the new version of MAID, the $S$, $P$, $D$ and $F$ waves of the background
contributions are unitarized in accordance with the K-matrix approximation,
\begin{equation}
 t_{\gamma\pi}^{B,\alpha}({\rm MAID})=
 \exp{(i\delta^{\alpha})}\,\cos{\delta^{\alpha}}
 v_{\gamma\pi}^{B,\alpha}(W,Q^2).
\label{eq6}
\end{equation}

From Eqs.~(\ref{eq3}) and (\ref{eq6}), one finds that the difference between the
background terms of MAID and of the dynamical model is that off-shell rescattering
contributions (principal value integral) are not included in MAID, therefore, after
re-fitting the data, they are implicitly contained in the resonance part leading to
dressed resonances.

Following  Ref.~\cite{Maid},  we assume a Breit-Wigner form for the resonance
contribution of the $e.m.$ multipoles ${\mathcal A}^{R}_{\alpha}(W,Q^2)$, which are the
the partial waves of the T-matrix $t_{\gamma\pi}^{R,\alpha}(W,Q^2)$ in Eq.~(\ref{eq5}),
\begin{equation}
{\mathcal A}_{\alpha}^R (W,Q^2)\,=\, \bar{\mathcal A}_{\alpha}^R (Q^2)\, \frac{f_{\gamma
R}(W)\Gamma_R(W)\,M_R\,f_{\pi R}(W)\,c_{\pi R}}{M_R^2-W^2-iM_R\Gamma_R(W)}
\,e^{i\phi(W)}, \label{eq7}
\end{equation}
where $f_{\pi R}$ is the usual Breit-Wigner factor describing the decay of a resonance
$R$ with total width $\Gamma_{R}(W)$ and physical mass $M_R$. The expressions for
$f_{\gamma R}, \, f_{\pi R}$ and $\Gamma_R$ are given in Ref.~\cite{Maid}. The factor
$c_{\pi R}$ is $\sqrt{3/2}$ and $-1/\sqrt{3}$ for resonances with isospin $3/2$ and
isospin $1/2$, respectively. The phase $\phi(W)$ in Eq.~(\ref{eq7}) is introduced to
adjust the phase of the total multipole to equal  the corresponding $\pi N$ phase shift
$\delta^{\alpha}$. While in the original version of MAID only the 7 most important
nucleon resonances were included with mostly only transverse e.m. couplings, from version
MAID2003 on all four star resonances below $W=2$~GeV are included. These are
$P_{33}(1232)$, $P_{11}(1440)$, $D_{13}(1520)$, $S_{11}(1535)$, $S_{31}(1620)$,
$S_{11}(1650)$, $D_{15}(1675)$, $F_{15}(1680)$, $D_{33}(1700)$, $P_{13}(1720)$,
$F_{35}(1905)$, $P_{31}(1910)$ and $F_{37}(1950)$.

The resonance couplings $\bar{\mathcal A}_{\alpha}^R(Q^2)$ can be taken as constants in a
single-Q$^2$ analysis, e.g. in photoproduction, where $Q^2=0$ but also at any fixed
$Q^2$, where enough data with W and $\theta$ variation is available. Alternatively they
can also be parametrized as functions of $Q^2$. Then it is possible to determine the
values $\bar{\mathcal A}_{\alpha}^R(0)$ from a fit to the world database of
photoproduction, while the parameters of the $Q^2$ evolution
can be obtained from a combined fitting of all electroproduction data at different $Q^2$.
The latter procedure we call the `superglobal fit'. In MAID the photon couplings
$\bar{\mathcal A}_{\alpha}^R$ are input parameters. They are directly related to the
helicity couplings $A_{1/2}, A_{3/2}$ and $S_{1/2}$ of nucleon resonance excitation as
described later in the text.

\section{Partial Wave Analysis}
\label{sec3}

The unitary isobar model MAID was used to analyze the world data of pion photo- and
electroproduction. In a first step we have fitted the background parameters of MAID and
the transverse normalization constants $\bar{\mathcal A}_{\alpha}^R (0)$ for the nucleon
resonance excitation. The latter ones give rise to the helicity couplings. Most of the
couplings are in good agreement with PDG and the GW/SAID analysis.

\begin{figure}[htp]
      \includegraphics[height=12cm, angle=90]{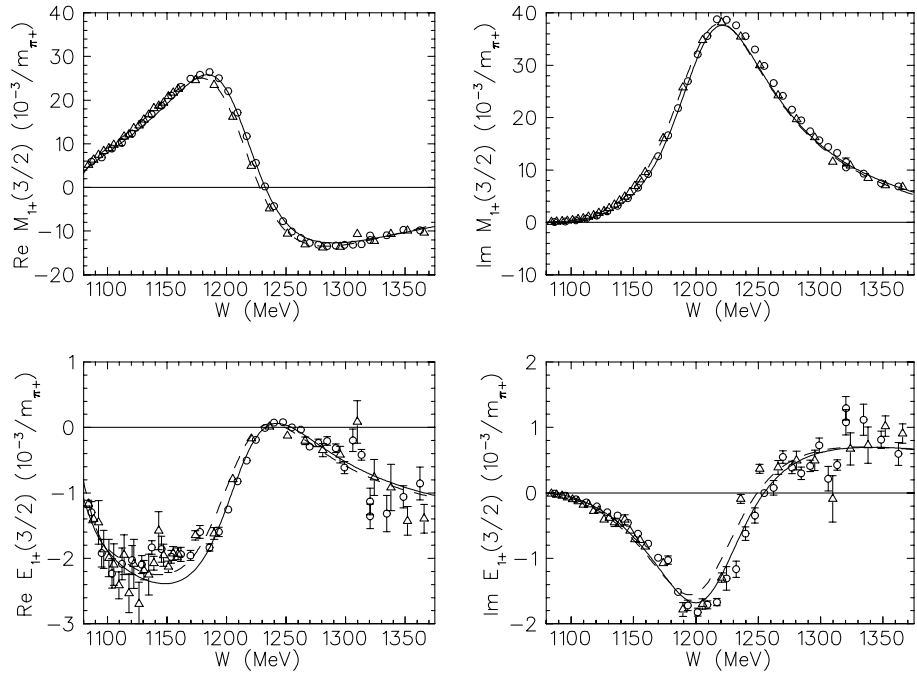}
      \caption{Real and imaginary parts of the $P_{33}$ multipoles $M_{1+}$ and
      $E_{1+}$. Our solution MAID05 (solid curve) is compared to the GW/SAID solution
      SM02 (dashed curve). The single-energy solutions (data points) are from MAID (open circles) and
      SAID (open triangles).} \label{fig1}
\end{figure}

\begin{figure}[htp]
      \includegraphics[height=12cm, angle=90]{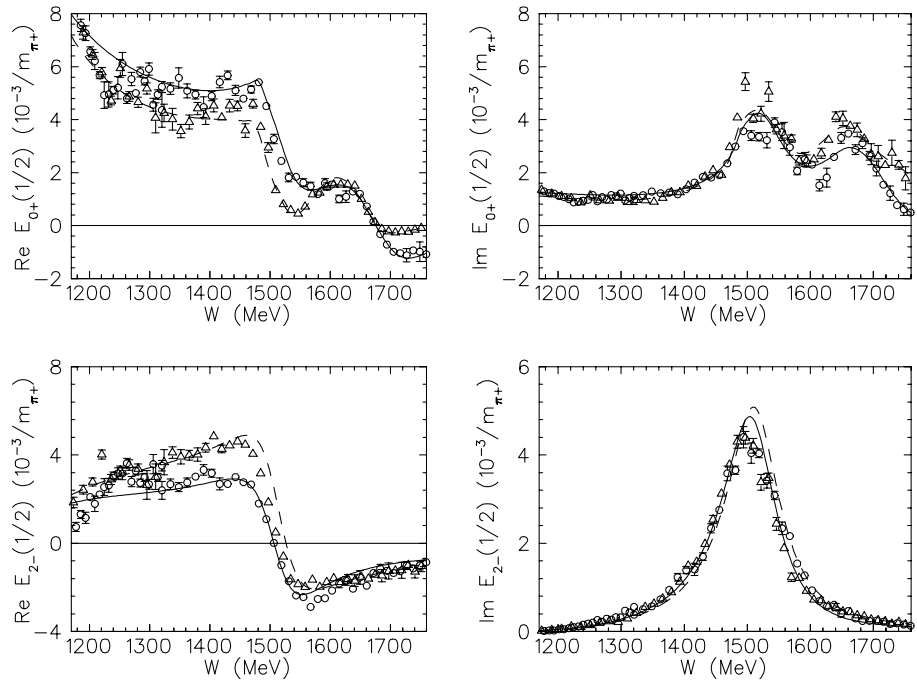}
      \caption{Real and imaginary parts of the $S_{11}$ multipole $E_{0+}$ and
      the $D_{13}$ mutipole $E_{2-}$. Notation as in Fig.~\ref{fig1}} \label{fig2}
\end{figure}

In Fig.~\ref{fig1} and Fig.~\ref{fig2} we give a comparison between MAID and SAID for
four important multipoles, $M_{1+}(P_{33}), E_{1+}(P_{33}), E_{0+}(S_{11})$ and
$E_{2-}(D_{13})$. For both analyses we show the global (energy dependent) curves together
with the local (single energy) fits, where only data in energy bins of 10-20 MeV are
fitted. Such a comparison demonstrates the fluctuations due to a limited data base or due
to a weak sensitivity of the available data to a small multipole, e.g. in the case of the
$E_{0+}$ or $E_{1+}$ multipoles. The very large efforts that had been made at MAMI and
LEGS in the 90s to precisely determine the $E/M$ ratio shows up with much better data
points for $E_{1+}$ around the Delta resonance position. Fig.~\ref{fig2}  also shows
systematic differences between the MAID and SAID analyses in the real parts of $E_{0+}$
and $E_{2-}$. Due to correlations between these amplitudes, the differences cannot be
resolved with our current data base. Because of isospin 1/2, they can, however, lead to
sizeable differences in the $\gamma,\pi^+$ channel, where the data base is still quite
limited.


In a second step we have fitted recent differential cross section data on $p(e,e'p)\pi^0$
from Mainz~\cite{Pos01}, Bates~\cite{Mer01}, Bonn~\cite{Ban02} and
JLab~\cite{Lav01,Smit03,Fro99}. These data cover a $Q^2$ range from $0.1\cdots 4.0$
GeV$^2$ and an energy range $1.1$ GeV $< W < 2.0$ GeV. In a first attempt we have fitted
each data set at a constant $Q^2$ value separately. This is similar to a partial wave
analysis of pion photoproduction and only requires additional longitudinal couplings for
all the resonances. The $Q^2$ evolution of the background is described with nucleon Sachs
form factors in the case of the $s-$ and $u-$ channel nucleon pole terms. At the e.m.
vertices of the $\pi$ pole and seagull terms we apply the pion and axial form factors,
respectively, while a standard dipole form factor is used for the vector meson exchange.
Furthermore, as mentioned above, we have introduced a $Q^2$ evolution of the transition
form factors of the nucleon to $N^*$ and $\Delta$ resonances and have parameterized each
of the transverse $A_{1/2}$ and $A_{3/2}$ and longitudinal $S_{1/2}$ helicity couplings.
In a combined fit with all electroproduction data from the world data base of
GWU/SAID~\cite{SAID} and the data of our single-$Q^2$ fit we obtained a $Q^2$ dependent
solution (superglobal fit).

\section{The Nucleon to Delta(1232) Transition}
\label{sec4}

For the resonance couplings of the $P_{33}$ multipoles we find a convenient
parametrization in the following form
\begin{equation}
\bar{\mathcal A}_{\alpha}^\Delta(Q^2) = \bar{\mathcal A}_{\alpha}^\Delta(0)
\frac{k(W,Q^2)}{k(W,0)} (1+\beta_\alpha Q^{n_\alpha})\, e^{-\gamma Q^2}\,
G_D(Q^2)\,\label{eq9}
\end{equation}
with the usual dipole form factor $G_D(Q^2)=1/(1+Q^2/0.71GeV^2)^2$. Even though the
exponential does not lead to the proper asymptotic behaviour of the helicity amplitudes,
by this ansatz we get a much simpler parametrization which is well behaved in the
physical region. On the other hand, the asymptotic behavior as predicted by pQCD is yet
far outside of the currently accessible experimental range. The fitted parameters are
listed in Table~\ref{tab1}.

\begin{table}[htbp]
\caption{Parametrization of the $Q^2$ dependence for the electromagnetic $N \rightarrow \Delta$
amplitudes. All 3 amplitudes have the same exponential factor with
$\gamma=0.21$~GeV$^{-2}$ and $G_D(Q^2)$ is the dipole form factor
$1/(1+Q^2/0.71GeV^2)^2$. } \label{tab1}
\begin{tabular}{c|ccc}
\hline  & $\bar{\mathcal A}_\alpha^\Delta(0)$ ($10^{-3}$ GeV$^{-1/2}$) & $n_\alpha$ & $\beta_\alpha$ (GeV$^{-n_\alpha}$)\\
\hline
M1  & 294.  &  2  &   0   \\
E2  & -6.37 &  2  & -0.31 \\
C2  & -19.5 &  6  & 0.0167\\
\hline
\end{tabular}
\end{table}

With the definition of Eq.~(\ref{eq6}), the background amplitudes of the $P_{33}$ channel
vanish exactly at the resonance position and the resonance amplitudes become purely
imaginary. In this case the helicity amplitudes $A_{1/2}, A_{3/2}, S_{1/2}$, which are
the characteristic numbers for $e.m.$ resonance excitation, are directly related to the
resonance multipoles at $W=M_\Delta$ (see e.g. PDG94 or Ref.~\cite{Arn90}),
\begin{eqnarray}
A_{1/2} &=& -\frac{1}{2} (\bar{M}_{1+}^{(3/2)} + 3 \bar{E}_{1+}^{(3/2)})\,,\nonumber\\
A_{3/2} &=& -\frac{\sqrt{3}}{2} (\bar{M}_{1+}^{(3/2)} - \bar{E}_{1+}^{(3/2)})\,,\nonumber\\
S_{1/2} &=& -\sqrt{2} \bar{S}_{1+}^{(3/2)}\,\nonumber\\
\mbox{with}\;\;
\{\bar{M},\bar{E},\bar{S}\}_{1+}^{(3/2)}&=&\left( \frac{8\,\pi\, q_\Delta\, M_\Delta\,
\Gamma_\Delta}{3\, k_W\, M_N } \right)^{1/2}\mbox{Im}
\{M,E,S\}_{1+}^{(3/2)}(W=M_\Delta)\,.\label{eq10}
\end{eqnarray}
In MAID these reduced resonance multipoles are identical to $\bar{\cal
A}_{\alpha}^{\Delta}(Q^2)$. Using this definition we can now also define the
$N\rightarrow\Delta$ transition form factors
\begin{eqnarray}
G_M^*(Q^2) &=& \,\,\, b_\Delta \bar{M}_{1+}^{(3/2)}(Q^2)
=b_\Delta \bar{{\cal A}}_{M}^{\Delta}(Q^2)\,,\nonumber\\
G_E^*(Q^2) &=& -b_\Delta \bar{E}_{1+}^{(3/2)}(Q^2)
=-b_\Delta \bar{\cal A}_{E}^{\Delta}(Q^2)\,,\nonumber\\
G_C^*(Q^2) &=& - b_\Delta \frac{2M_\Delta}{k_\Delta}\bar{S}_{1+}^{(3/2)}(Q^2) = -
b_\Delta \frac{2M_\Delta}{k_\Delta} \bar{{\cal A}}_{C}^{\Delta}(Q^2)\,\nonumber\\
\mbox{with}\;\; b_\Delta &=& \left( \frac{8\, M_N^2\, q_\Delta\,
\Gamma_\Delta}{3\,\alpha_{em}\, k_\Delta^2} \right)^{1/2} \,,\label{eq11}
\end{eqnarray}
$\alpha_{em}=1/137$, $\Gamma_{\Delta}=115$ MeV and $k_{\Delta}$, $q_{\Delta}$ the photon
and pion momenta at $W=M_\Delta\,$. Eq.~(\ref{eq11}) corresponds to the definition of Ash
\cite{Ash67}, the form factors of Jones and Scadron \cite{Jon73} are obtained by
multiplication with an additional factor, $G^{JS}=\sqrt{1+Q^2/(M_N +M_\Delta)^2} \,
G^{Ash}$.

The $e.m.$ transition form factors may also be expressed in terms of the helicity
amplitudes $A_{1/2}, A_{3/2}$ and $S_{1/2}$, which are determined at the resonance
position $W=M_\Delta$ and are functions of $Q^2$,
\begin{eqnarray}
G_M^* &=&  -c_\Delta
           (A_{1/2}+\sqrt{3} A_{3/2})\,,\nonumber\\
G_E^* &=&  \,\,\, c_\Delta
           (A_{1/2}-\frac{1}{\sqrt{3}} A_{3/2})\,,\nonumber\\
G_C^* &=&
\sqrt{2} c_\Delta \frac{2M_\Delta}{k_\Delta}  S_{1/2}\,,\nonumber\\
\mbox{with} \;\; c_\Delta &=& \left( \frac{m^3 k_W}{4\pi\alpha M_\Delta k_\Delta^2}
 \right)^{1/2}
\end{eqnarray}
and the equivalent photon energy $k_W=k(W,Q^2=0)$. The $E/M$ and $S/M$ ratios can then be defined as
\begin{eqnarray}
R_{EM} &=& -\frac{G_E^*}{G_M^*}=\frac{A_{1/2}-\frac{1}{\sqrt{3}}
A_{3/2}}{A_{1/2}+\sqrt{3} A_{3/2}}\,,\nonumber\\
R_{SM} &=& - \frac{k_\Delta}{2M_\Delta}  \frac{G_C^*}{G_M^*}=\frac{\sqrt{2} S_{1/2}}{A_{1/2}+\sqrt{3} A_{3/2}}\,.
\end{eqnarray}
Here we use a definition for the $G_C^*$ form factor, which is now mostly used in the
literature. It differs from our previous definition~\cite{Tiat03} by a factor of
$k_\Delta/2M_\Delta$.

\begin{figure}[htbp]
      \includegraphics[height=13cm, angle=90]{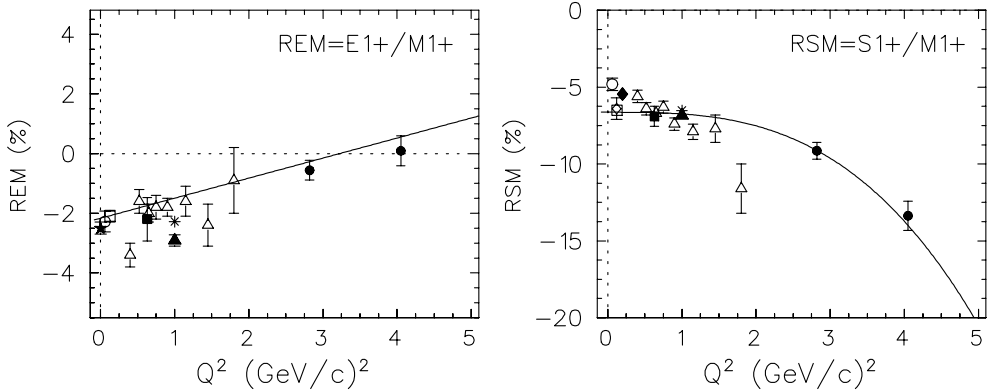}
      \caption{$R_{EM}$ and $R_{SM}$ ratios of the $\Delta(1232)$ resonance excitation
      at large $Q^2$. The solid curves are the MAID2005 solutions.
      The experimental analyses are from: MAMI(solid star\cite{Beck97}, open
      circles\cite{Stav06}, open diamond\cite{Pos01}, solid diamond~\cite{Elsn06}),
      Bates(open squares\cite{Mer01}), ELSA(solid squares\cite{Ban02}), JLab(open
      triangles\cite{Joo02}, solid triangles\cite{Kell05a}). The solid circles and asterisks
      show our own analysis to the JLab Hall C data\cite{Fro99} and JLab Hall A
      data\cite{Kell05}, respectively.
       } \label{fig3}
\end{figure}

\begin{figure}[htbp]
      \includegraphics[height=13cm, angle=90]{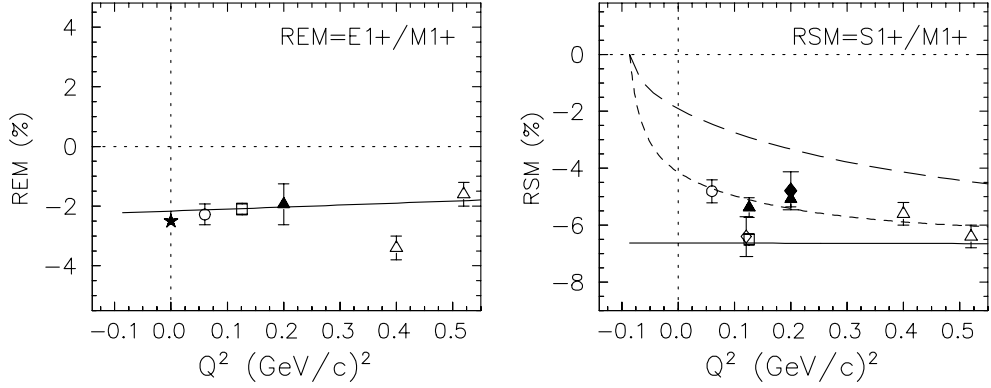}
      \caption{$R_{EM}$ and $R_{SM}$ ratios of the $\Delta$ excitation
      at low $Q^2$. The solid triangles show a preliminary analysis with MAMI data at
      $Q^2=0.2$~GeV$^2$ and a combined analysis of Bates and MAMI data at
      $Q^2=0.127$~GeV$^2$~\cite{Spar06}, all other experimental analyses are as in the previous figure.
      The solid curves are the MAID2005 solutions, whereas the dashed curves
      show two solutions that are consistent with the Siegert theorem at
      pseudothreshold, see text.} \label{fig4}
\end{figure}

In Fig.~\ref{fig3} we show our MAID2005 solution which can also be evaluated with the
definitions of Eq.~(\ref{eq9}) and numerical numbers given in Table~\ref{tab1}. The
curves are compared to $R_{EM}$ and $R_{SM}$ ratios determined in different experimental
analyses from MAMI, ELSA, Bates and JLab. The points at the two largest $Q^2$ values of
2.8 and 4.0 GeV$^2$ have been re-analyzed by our group~\cite{DMT01}. Especially our
$R_{EM}$ values are more positive than in the original experimental analysis of Frolov et
al.~\cite{Fro99}, which was done in a truncated multipole expansion. Also at
$Q^2=1.0$~GeV$^2$ we have performed such a re-analysis with the data of Kelly et
al.~\cite{Kell05} and also find an $E/M$ ratio closer to our global solution. In this
case, however, the data set is almost complete and consists out of 16 different
polarization observables. But, of course also for this data set the angular resolution is
limited and the observables have sizable errors in some cases. Therefore, we also find
for this analysis a model uncertainty due to higher partial waves.
In Fig.~\ref{fig4} we have enlarged the low $Q^2$ region, where the most recent
experimental analysis from MAMI and Bates are shown. We note that in this region also
preliminary data of CLAS have been analyzed and reported elsewhere in this
proceedings~\cite{Smit06}. Our global MAID solution which is practically constant in this
$Q^2$ region is completely consistent with the individual data points of $R_{EM}$ but
does not have the right shape for $R_{SM}$. It was reported many times previously that in
dynamical models $R_{SM}$ has a tendency to rise and this was attributed to the pion
cloud effects. In a dynamical calculation such a behavior arises naturally, in MAID,
where the transition amplitudes are empirically parameterized, it has to be included by
hand.
As pointed out in more detail in Ref.~\cite{Dre06} an important property is the
model-independent behavior of the multipoles at physical threshold (vanishing pion
momentum, $q= |\vec q|=0)$ and at pseudothreshold (Siegert limit, vanishing photon
momentum $k= |\vec k|=0)$. In particular we find the following conditions for the Delta
multipoles:
$(E_{1+}, M_{1+}, L_{1+}) \rightarrow k q$.
Due to gauge invariance, $k L_{1+}=\omega S_{1+}$, the Coulomb amplitudes acquire an
additional factor k at pseudothreshold, i.e., $S_{1+} \sim k L_{1+}$. The Siegert limit
is reached at $Q^2=Q^2_{pt}=-(W-M_N)^2$, and since $\vec k = 0$ means that no direction
is defined, the electric and longitudinal multipoles have to be equal at this point,
$L_{1+}(Q_{pt}^2)=E_{1+}(Q_{pt}^2)$ and $S_{1+}(Q_{pt}^2)=k/\omega\, E_{1+}(Q_{pt}^2)$.
For the ratios at $W=M_\Delta$, we do not get a further constraint for $R_{EM}$, it
remains finite at $Q^2_{pt}$, but for $R_{SM}$ we get the important condition
\begin{equation}
R_{SM}=\frac{S_{1+}}{M_{1+}}=\frac{k_\Delta}{\omega_\Delta}\frac{L_{1+}}{M_{1+}}
=\frac{k_\Delta}{\omega_\Delta}\frac{E_{1+}}{M_{1+}}=\frac{k_\Delta}{M_\Delta-M_N}R_{EM},\quad
\mbox{for} \quad k_\Delta\rightarrow 0\,.
\end{equation}
This Siegert approximation is shown as the long-dashed line in Fig.~\ref{fig4}. In this
simplest form it looks very similar to the dynamical calculation of Sato and
Lee~\cite{SL96,SL01}. With further moderate adjustments in the $Q^2$ shape the empirical
MAID parametrization can be improved to get a perfect fit through the data points
(short-dashed line). However, this preliminary result is only a fit of the single-Q2
points and not yet a fit to the measured cross sections.


In a recent paper of Elsner et al.~\cite{Elsn06} we have analyzed the
longitudinal-transverse asymmetries $LT$ and $LT'$ measured at Mainz in two different
experiments at $W=1232$~MeV and $Q^2=0.2$~GeV$^2$. These asymmetries are related to the
partial cross sections via
\begin{eqnarray}
  \rho_{LT}(\theta) =
  \frac{\sqrt{2\epsilon(1+\epsilon)}\,d\sigma_{LT}}{d\sigma_{T}+\epsilon\,d\sigma_{L}+\epsilon\,d\sigma_{TT}}
  \quad\mbox{and}\quad
  \rho_{LT'}(\theta,\phi) =
  \frac{\sqrt{2\epsilon(1-\epsilon)}\,d\sigma_{LT'}\sin\phi}{d\sigma_{T}+\epsilon\,d\sigma_{L}
  +\epsilon\,d\sigma_{LT} \cos\phi + \epsilon\,d\sigma_{TT}\cos{2\phi}}\,.
  \label{eq15}
\end{eqnarray}
The sensitivity to the leading multipoles $M_{1+}$, $S_{+}$ and $S_ {0+}$ is shown by a
partial wave decomposition, where only the leading multipoles are retained,
\begin{eqnarray}
  \label{eq16}
  \rho_{LT}(\theta) &\simeq&
  f1(\theta)\frac{\Re\{(S_{0+} + 6  S_{1+}
  \cos\theta)M_{1+}^{\ast}\}}{|M_{1+}|^{2}}\,,\nonumber\\
  \rho_{LT'}(\theta,\phi) &\simeq&
  f2(\theta,\phi)\sin\phi \frac{\Im\{(S_{0+} + 6  S_{1+}
  \cos\theta)M_{1+}^{\ast}\}}{|M_{1+}|^{2}}\,.
\end{eqnarray}
At the resonance position, the $P_{33}$ multipoles become purely imaginary, therefore,
$\rho_{LT}$ which is proportional to the real part of the product becomes very sensitive
to the Delta resonance, whereas the imaginary part of the bilinear products in
$\rho_{LT'}$ show up with an enhanced sensitivity to the background multipoles. Such
background multipoles as $S_{0+}$ for isospin 1/2 and 3/2 and $S_{1+}$ for isospin 1/2
are not so well described within the unitary isobar model MAID. Only via PS-PV mixing,
the vector meson contributions and the unitarization procedure they can get non-Born
contributions. However, all these parameters are practically fixed, therefore we have
allowed for an additional variation of a real and imaginary non-Born part in the $S_{0+}$
multipole.

\begin{figure}[htbp]
      \includegraphics[height=5cm]{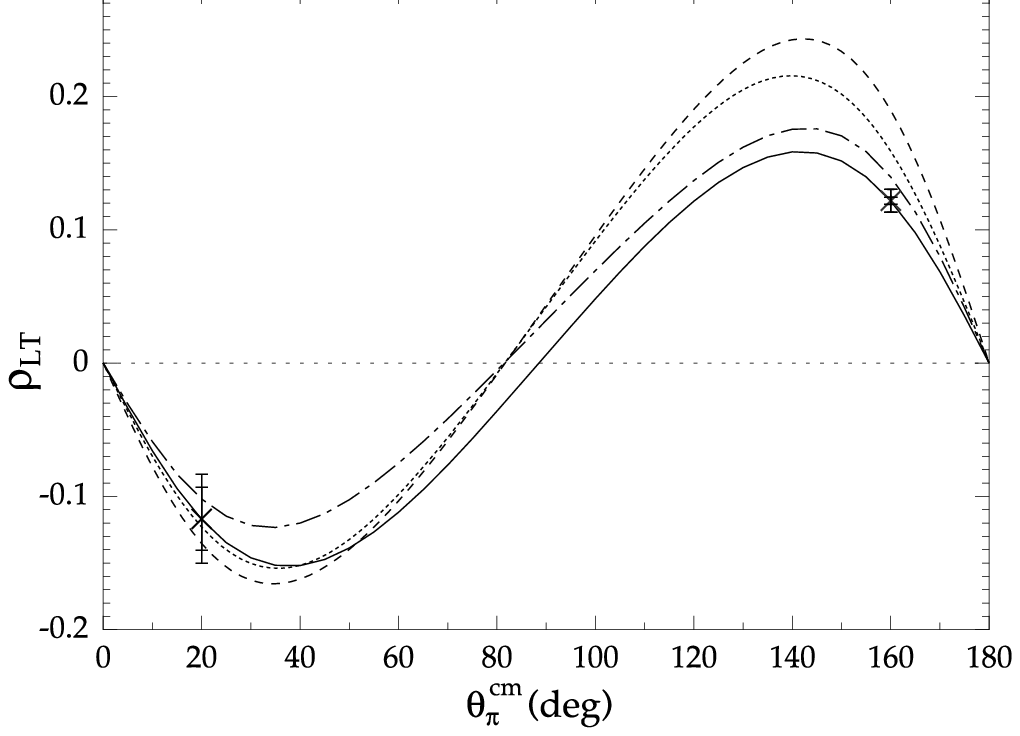}\hspace*{0.5cm}
      \includegraphics[height=5cm]{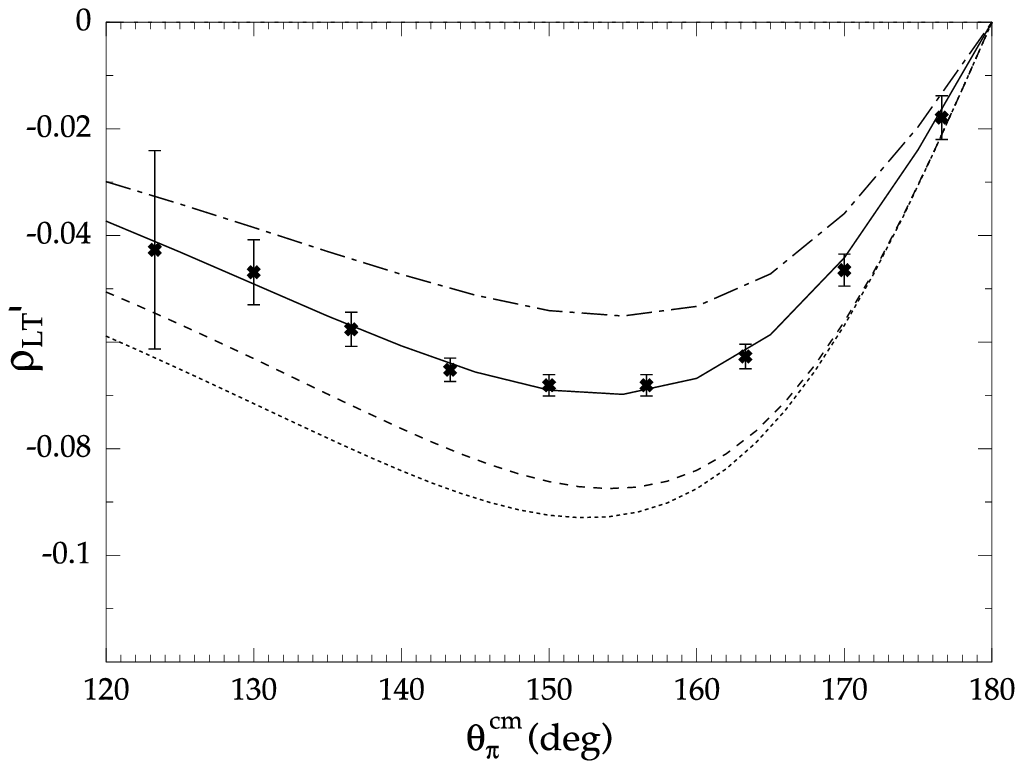}
      \caption{Left: $\rho_{LT}$ asymmetries compared with model predictions from MAID2003 \cite{Maid} (dotted),
  DMT2001 \cite{KY99,DMT01} (dashed), Sato/Lee
  \cite{SL01} (dashed dotted). The full curve represents the MAID2003 re-fit.
  Right: $\rho_{LT'}$ asymmetries from reference \cite{Bart02} with model
  predictions from MAID2003 \cite{Maid} (dotted),
  DMT2001 \cite{KY99,DMT01} (dashed), Sato/Lee \cite{SL01} (dashed dotted).
  The full curve represents the MAID2003 re-fit.  } \label{fig5}
\end{figure}

Using the new $\rho_{LT}$ data of Elsner~\cite{Elsn06} in conjunction with the previous
measurement of the $\rho_{LT\prime}$ asymmetry  of Bartsch et al. \cite{Bart02}, we
performed a re-fit of the MAID2003 parameters. We obtained sensitivity to real and
imaginary parts of the $S_{1+}$ and $S_{0+}$ amplitudes in the $p\pi^0$ channel. The results
for $\rho_{LT}$ and $\rho_{LT\prime}$ are depicted in Fig. \ref{fig5}, in comparison with
the standard MAID2003 and the dynamical models DMT2001 \cite{KY99,DMT01} and SL2001 \cite{SL01}.
Our MAID2003 re-fit results are also compared with a truncated analysis of
Eq.~(\ref{eq16}) using only the 3 partial waves $S_{1+}$, $S_{0+}$ and $M_{1+}$ and with
previous calculations in Table \ref{tab2} as multipole ratios for the $(p \pi^0)$
channel. The denoted errors are due to the re-fit of $S_{1+}$ and $S_{0+}$ within the
MAID2003 analysis taking into account the statistical and systematical errors. The model
dependence of the extraction can be estimated from the truncated multipole result given
in the second row in the table.
\begin{table}[h]
  \caption{Comparison of multipole ratios from asymmetry data and calculations, as
  discussed in the text.} \label{tab2}
  \begin{tabular}{lllll}
    \hline
    & $\frac{\Re\{S_{1+}M_{1+}^{\ast}\}}{{|M_{1+}|^{2}}}$ (\%)
    & $\frac{\Im\{S_{1+}M_{1+}^{\ast}\}}{{|M_{1+}|^{2}}}$ (\%)
    & $\frac{\Re\{S_{0+}M_{1+}^{\ast}\}}{{|M_{1+}|^{2}}}$ (\%)
    & $\frac{\Im\{S_{0+}M_{1+}^{\ast}\}}{{|M_{1+}|^{2}}}$ (\%)\\
    \hline
    \hline
MAID2003 re-fit & -5.45$\pm$0.42 & -2.92$\pm$0.48 & 2.56$\pm$2.25 & 3.67$\pm$2.33 \\

from Eq.~(\ref{eq16})& -4.78$\pm$0.69 & -- & 0.56$\pm$3.89 & --\\

MAID2003 &  -6.65   & -2.22  & 7.98   & 15.0  \\
DMT 2001 &  -6.81   & -2.20  & 7.33   & 12.4  \\

SL 2001  &  -4.74    & -1.77 & 5.14  & 4.35\\
    \hline
  \end{tabular}
\end{table}

A comparison between the MAID2003 re-fit and the truncated result shows agreement for the
$S/M$ ratio (first column of Table 2). In general, however, the truncation can lead to
wrong results as for the real part of the $S_{0+}/M_{1+}$ ratio, but the large error
takes account of this problem. The truncation in $\rho_{LT'}$ is even worse and would
lead to completely wrong results for the imaginary parts of the ratios, therefore, we
have omitted these numbers in the table. The biggest effect of our re-fit is observed in
the imaginary part of $S_{0+}/M_{1+}$, which is reduced by a factor of 4 compared to the
standard MAID calculation. This result shows the importance of non-Born terms in the
background amplitude $S_{0+}$ due to the pion cloud effects. Our fitted result is in good
agreement with the dynamical model of Sato/Lee but in disagreement with the DMT model.

\section{Photocoupling Amplitudes}
\label{sec5}

Besides our extensive studies of the $\Delta(1232)$ resonance, we have also investigated
the $Q^2$ evolution of all other resonance excitations that are included in MAID. At
higher energies, above the $\Delta$ region, however, the data are no longer so abundant
as in the $\Delta$ region, but due to a large data set of the CLAS
collaboration~\cite{Joo02,Smit03} for $p(e,e'\pi^0)p$ we were able to determine the
transverse and longitudinal helicity photon couplings as functions of $Q^2$ for all
4-star resonances below 1700~MeV. The data are available in the kinematical region of
$1100~\mbox{MeV}<W<1680~\mbox{MeV}$ and $0.4~\mbox{GeV}^2<Q^2<1.8~\mbox{GeV}^2$.

\begin{figure}[ht]
      \includegraphics[height=12.5cm, angle=90]{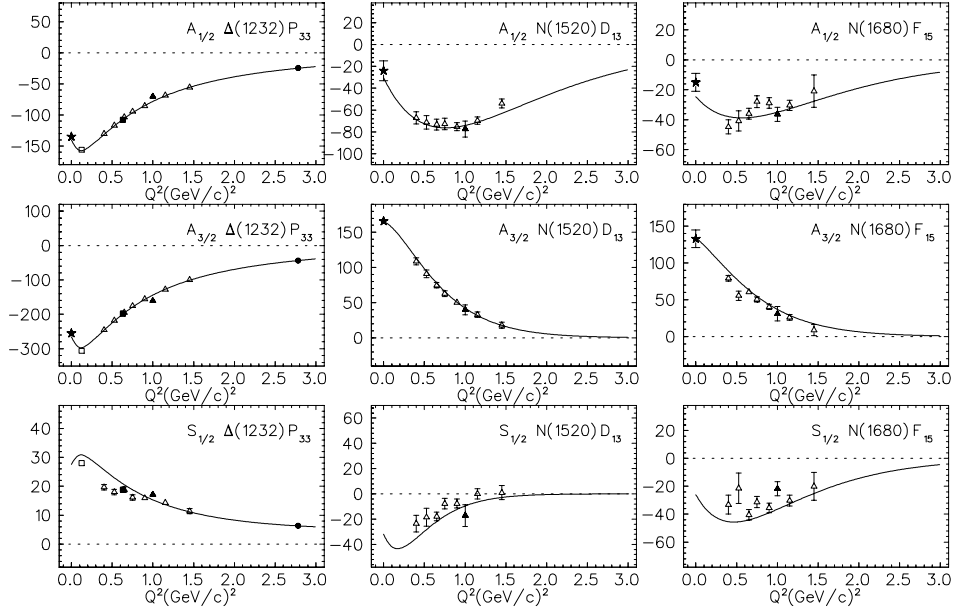}
      \caption{The $Q^2$
dependence of the transverse ($A_{1/2},A_{3/2}$) and longitudinal ($S_{1/2}$) helicity
couplings for the $P_{33}(1232)$, $D_{13}(1520)$ and $F_{15}(1680)$ resonance excitation.
The solid curves show our superglobal fit. The data points at finite $Q^2$ are obtained
from our single-$Q^2$ analysis using the data from MAMI and Bates for $Q^2=0.1$GeV$^2$,
from ELSA for 0.6 GeV$^2$, JLAB(Hall A) for 1.0 GeV$^2$, JLab(Hall C) for 2.8 GeV$^2$ and
JLab(Hall B) for the remaining points. At the photon point ($Q^2=0$) we show our result
of the photoproduction analysis.} \label{fig6}
\end{figure}

Above the $2\pi$ threshold the two-channel unitarity and consequently the Watson theorem
no longer hold. Therefore, the background amplitude (Eq.~(\ref{eq6})) of any partial wave
$\alpha$ does not vanish at the resonance position, except for the $P_{33}(1232)$. This
leads naturally to a model dependence in the resonance-background separation. Here in our
analysis we use the concept of the MAID model, where  the background and resonance
contributions are already separated according to the K-matrix approximation,
Eqs.~(\ref{eq5},\ref{eq6}). Therefore, we can start with Eq.~(\ref{eq7}) for a general
definition of the photon coupling amplitudes (see also PDG94 or Ref.~\cite{Arn90}). At
the resonance position, $W=M_R$ we then obtain
\begin{eqnarray}
{\cal A}_{\alpha}^R(M_R,Q^2)\, &=& i\,{\bar{\cal A}}_{\alpha}^R(Q^2) f_{\gamma R}(M_R)
f_{\pi R}(M_R)
c_{\pi R} e^{i \phi(M_R)}\nonumber\\
&=& {\cal A}_{\alpha}^{res}(M_R,Q^2)\, e^{i \phi(M_R)}\label{eq17}
\end{eqnarray}
with $f_{\gamma R}(M_R)=1$ and
\begin{equation}
f_{\pi R}(M_R)=\left[ \frac{1}{(2j+1)\pi} \frac{k_W}{|q|} \frac{M_N}{M_R}
\frac{\Gamma_{\pi N}}{\Gamma^2_{tot}} \right]^{1/2}\,.
\end{equation}
The factor $c_{\pi R}$ is $\sqrt{3/2}$ and $-1/\sqrt{3}$ for the isospin $3/2$ and
isospin $1/2$ multipoles, respectively. This leads to the definition
\begin{equation}
\bar{\cal A}_{\alpha}^R(Q^2) = \frac{1}{c_{\pi R} f_{\pi R}(M_R)} \mbox{Im} {\cal
A}_{\alpha}^{res} (M_R, Q^2)\,.
\end{equation}
It is important to note that by this definition the phase factor $e^{i\phi}$ in Eqs.
(\ref{eq7}) and (\ref{eq17}) is not part of the resonant amplitude but rather part of the
unitarization procedure. In the case of the $\Delta(1232)$ resonance this phase vanishes
at the resonance position due to Watson's theorem, however, for all other resonances it
is finite and in some extreme cases it can reach values of about $60^0$. $\bar{\cal
A}_{\alpha}^R$ is a short-hand notation for the electric, magnetic and longitudinal
multipole photon couplings of a given partial wave $\alpha$. As an example, for the
$P_{33}$ partial wave the specific couplings are denoted by $\bar{E}_{1+}$,
$\bar{M}_{1+}$ and $\bar{S}_{1+}$. By linear combinations they are connected with the
more commonly used helicity photon couplings $A_{1/2}$, $A_{3/2}$ and $S_{1/2}$. For
resonances with total spin $j=\ell + 1/2$ we get
\begin{eqnarray}
A^{\ell +}_{1/2} &=& -\frac{1}{2} [(\ell +2) \bar{E}_{\ell+} + \ell \bar{M}_{\ell+} ]\,, \nonumber\\
A^{\ell +}_{3/2} &=& \frac{1}{2}\sqrt{\ell(\ell+2)} (\bar{E}_{\ell+} - \bar{M}_{\ell+})\,, \nonumber\\
S^{\ell +}_{1/2} &=& -\frac{\ell+1}{\sqrt{2}} \bar{S}_{\ell+}
\end{eqnarray}
and for $j=(\ell+1) - 1/2$
\begin{eqnarray}
A^{(\ell+1)-}_{1/2} &=& \frac{1}{2} [(\ell +2) \bar{M}_{(\ell+1)-} - \ell \bar{E}_{(\ell+1)-} ]\,, \nonumber\\
A^{(\ell+1)-}_{3/2} &=& -\frac{1}{2}\sqrt{\ell(\ell+2)} (\bar{E}_{(\ell+1)-} + \bar{M}_{(\ell+1)-})\,, \nonumber\\
S^{(\ell+1)-}_{1/2} &=& -\frac{\ell+1}{\sqrt{2}} \bar{S}_{(\ell+1)-}\,.
\end{eqnarray}
In Fig.~\ref{fig6} we show our results for the $\Delta(1232)$, the $D_{13}(1520)$ and the
$F_{15}(1680)$ resonances. Our superglobal fit agrees very well with our single-$Q^2$
fits, except in the case of the $\Delta$ resonance for the 2 lowest points of $S_{1/2}$
from our analysis of the Hall B data. As we discussed before, this is due to the wrong
Siegert limit in MAID2005 for the longitudinal couplings, which will be corrected in the
next version of MAID. It will also have consequences for higher resonances, however, the
pseudothreshold at resonance energy moves further away from the physical region (e.g.
$Q^2_{pt}=-0.338$~GeV$^2$ for the $D_{13}(1520)$) and will not have such a large
consequence as in the case of the $\Delta(1232)$.

\begin{figure}[ht]
      \includegraphics[width=9cm]{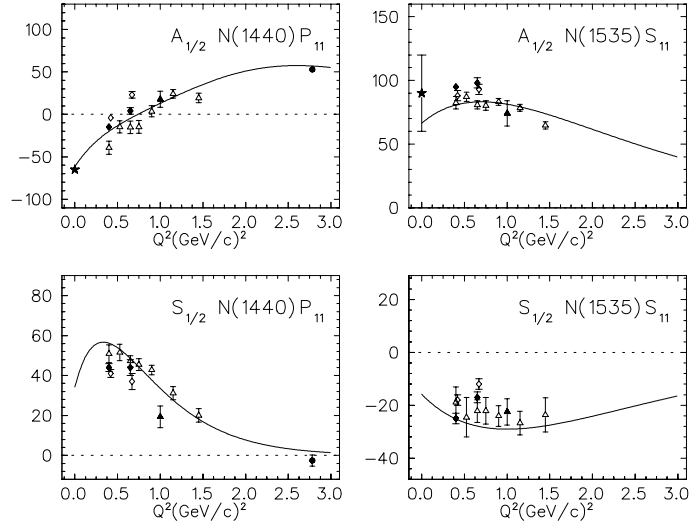}
     \caption{The $Q^2$ dependence of the transverse and longitudinal helicity amplitudes
for the $P_{11}(1440)$ and the $S_{11}(1535)$ resonance excitation of the proton. The
solid lines are the superglobal Maid2005 solutions. The solid red (gray) points are our
single-$Q^2$ fits to the exp. data from CLAS/JLab~\protect\cite{Smit03}, the solid and
open blue circles show the isobar and dispersion analysis of Aznauryan
\protect\cite{Azna05} using a similar data set.}
 \label{fig7}
\end{figure}
Our results for the spin 1/2 resonances $P_{11}(1440)$ and $S_{11}(1535)$ are shown in
Fig.~\ref{fig7}. As in the previous figure, our superglobal solutions are in a generally
good agreement with our single-$Q^2$ points (triangles and circles). In this figure we
also compare with the results of Aznauryan~\cite{Azna05} using a similar data set of CLAS
data. Both of our analyses are in good agreement, and furthermore, the overall
fluctuation of the points give a realistic estimate of the model uncertainty. For the
Roper resonance we notice a zero crossing of the transverse helicity coupling around
$Q^2\approx 0.7$~GeV$^2$ and a maximum at a relatively large momentum transfer of
2.5~GeV$^2$. Also the longitudinal Roper excitation exhibits very large values around
$Q^2\approx 0.5$~GeV$^2$, in fact this is the strongest longitudinal resonance excitation
that we can find. This should be a clear answer to the question raised by Li and
Burkert~\cite{Burk92}, whether the Roper resonance is a radially excited 3-quark state or
a quark-gluon hybrid state. In the latter case it was argued that the longitudinal
coupling would completely vanish. Also the couplings of the $S_{11}(1535)$ resonance are
quite strong. As it is already known form $\eta$ electroproduction, the transverse form
factor falls off very weakly. At $Q^2\approx 3$~GeV$^2$ it is much stronger than the
Delta or the $D_{13}$ and is comparable to the Roper. However, due to the much smaller
width of the $S_{11}$ compared to the Roper, the resonant amplitude (see
Eq.~(\ref{eq17})) of the $S_{11}$ dominates at large $Q^2$. This result is in agreement
with the observation in the inclusive electroproduction cross section of the proton,
where at small momentum transfer the $\Delta(1232)$ dominates but at larger momentum
transfer the second resonance region takes over.

\section{Conclusions}
\label{sec6}

Using the world data base of pion photo- and electroproduction and recent data from
Mainz, Bonn, Bates and JLab we have made a first attempt to extract all longitudinal and
transverse helicity amplitudes of nucleon resonance excitation for four star resonances
below $W=1.7$~GeV. For this purpose we have extended our unitary isobar model MAID and
have parameterized the $Q^2$ dependence of the transition amplitudes. Comparisons between
single-$Q^2$ fits and a $Q^2$ dependent superglobal fit give us confidence in the
determination of the helicity couplings of the $P_{33}(1232), P_{11}(1440), S_{11}(1535),
D_{13}(1520)$ and the $F_{15}(1680)$ resonances, even though the model uncertainty of
these amplitudes can be as large as 50\% for the longitudinal amplitudes of the $D_{13}$
and $F_{15}$.

For other resonances the situation is more uncertain. However, this only reflects the
fact that precise data in a large kinematical range are absolutely necessary. In some
cases double polarization experiments are very helpful as has already been shown in pion
photoproduction. Furthermore, without charged pion electroproduction, some ambiguities
between partial waves that differ only in isospin as $S_{11}$ and $S_{31}$ cannot be
resolved without additional assumptions. While all electroproduction results discussed
here are only for the proton target, we have also started an analysis for the neutron,
where much less data are available from the world data base and no new data has been
analyzed in recent years. Since we can very well rely on isospin symmetry, only the
electromagnetic couplings of the neutron resonances with isospin $1/2$ have to be
determined. We have obtained a superglobal solution for the neutron which is implemented
in MAID2005. However, for most resonances this is still highly uncertain. So it will be a
challenge for the experiment to investigate also the neutron resonances in the near
future.


\begin{theacknowledgments}
 This work was supported in part by the Deutsche Forschungsgemeinschaft (SFB443).
\end{theacknowledgments}

\end{document}